\documentstyle[preprint,aps,prb,floats,epsfig]{revtex}
\tightenlines
\begin{document}
\title{Grain boundary effects on magnetotransport\\
 in bi-epitaxial films of La$_{0.7}$Sr$_{0.3}$MnO$_3$.}

\author{R. Mathieu, P. Svedlindh}
\address{Department of Materials Science, Uppsala University\\
        Box 534, SE-751 21 Uppsala, Sweden}

\author{R. A. Chakalov\thanks{Current address: Dept. of Physics and
Astronomy, University of Birmingham, Edgbaston, Birmingham B15 2TT,
UK.}, Z. G. Ivanov}

\address{Department of Physics, Chalmers University of Technology,
SE-412 96 G{\"o}teborg, Sweden}

\date{\today}

\maketitle

\begin{abstract}
The low field magnetotransport of La$_{0.7}$Sr$_{0.3}$MnO$_3$ (LSMO)
films
grown on SrTiO$_3$ substrates has been investigated. A high qualtity
LSMO
film
exhibits anisotropic magnetoresistance (AMR) and a peak in the
magnetoresistance close to the Curie temperature of LSMO.
Bi-epitaxial films prepared using a seed layer of MgO and a buffer
layer of
CeO$_2$ display a resistance dominated by grain boundaries. One film
was
prepared with seed and buffer layers intact, while a second sample
was prepared as a 2D square array of grain boundaries. These films
exhibit
i) a low temperature tail in the low field magnetoresistance; ii) a
magnetoconductance with a
constant high field slope; and iii) a comparably large AMR effect. A
model
based on a two-step tunneling process, including spin-flip tunneling,
is
discussed and shown to be consistent with the experimental findings
of the
bi-epitaxial films.
\end{abstract}
\pacs{72.15.Gd, 75.30.Kz, 75.50.Cc}

\section{Introduction}
In recent years much attention has focused on the magnetoresistive
properties of hole doped manganite perovskites\cite{helmont}. In case
of single crystals\cite{hwang1} and high quality epitaxial
films\cite{hund1,donnell1,steenbeck12,shreekala,ziese12}, the
magnetoresistance effect is large only close to the ferromagnetic
transition temperature. Moreover, comparably large applied fields, of
order 1 T, are required to obtain a sizeable effect, which makes it
difficult to exploit these materials in for instance sensor applications.
It has been suggested that transport is of activated form with a hopping
motion of carriers forming polarons. Also, a strong transport-magnetism
correlation has been observed both
above and below the Curie temperature\cite{hund1,donnell1}. \\
A large low field magnetoresistance is known to exist in
polycrystalline
bulk ceramic materials\cite{balcells,lee} as well as in thin films
containing interfaces and grain boundaries of some
kind\cite{hwang1,steenbeck12,shreekala,ziese12,li,isaac,wang12}.
Experimental realizations of the latter include polycrystalline
films\cite{hwang1,shreekala,li,sun}, films grown on bi-crystal
substrates with different grain boundary
angles\cite{steenbeck12,isaac,westerburg},
step-edge structures\cite{ziese12} and trilayer junction
structures\cite{li,sun}.
Models proposed to explain this low field effect include
spin-polarized tunneling
\cite{lee,gu,guinea,lyu1,lyu2,inoue}, spin dependent scattering at
grain boundaries/domain
walls\cite{wang12} and activated carrier transport in grain boundary
regions\cite{evetts}. \\
In this paper, we study the magnetic and magnetoresistive properties
of a 2D array (2DA) of weakly coupled LSMO islands. Its properties are
compared with two reference samples: an epitaxial LSMO film (EF) and a LSMO
film with irregular grain boundaries (GBF). Both 2DA and GBF were prepared
as bi-epitaxial films and exhibit a
similar and strong effect of grain boundaries on the magnetoresistive
behavior. Still, differences in behavior are observed when studying
the anisotropy of the magnetoresistance. Below magnetic saturation,
the anisotropic magnetoresistance in case of 2DA contains an extrinsic
contribution
from the geometry of the grain boundaries. It is argued that a model
based on a two-step inelastic tunneling process can account for the
magnetoresistive behavior of the two bi-epitaxial films.

\section{Samples and experiments}

Three La$_{0.7}$Sr$_{0.3}$MnO$_3$ (LSMO) (a$_{LSMO}$=3.82 \AA) thin
films grown on SrTiO$_3$ (STO) substrates (a$_{STO}$=3.905 \AA) have been
investigated: A high quality epitaxial film (EF) and
two bi-epitaxial films. Structural properties of the
films were checked with X-ray $\theta-2\theta$ and $\phi$ scans.
Details on the fabrication process and characterization are
described elsewhere\cite{zdravkapl}. The epitaxial film is highly
c-axis oriented and
only $[100]$ LSMO $\parallel$ $[100]$ STO in-plane orientation is
observed. The bi-epitaxial\cite{biepitax} films were prepared by
using a seed layer of MgO (a$_{MgO}$=4.21 \AA) having a thickness of
20 nm and a buffer layer of
CeO$_2$ (a$_{CeO}$=5.41 \AA). One film sample (GBF) was
prepared with seed and buffer layers intact, while a second sample
(2DA) was prepared as a 2D square array of grain boundaries. To form this
array, the MgO seed layer was etched into a chess board pattern with fields of
$8\times8 \mu$m$^2$. The chess board fields,
where the STO surface is disclosed, initiate a 45$^o$ in-plane rotated
growth of
the CeO$_2$ buffer layer. The LSMO inherits the template orientation
of the buffer layer forming 45$^o$ misoriented domains as well as a 500
$\times$ 500
array of 45$^o$ grain boundaries (GB). Fig.~\ref{fig0} shows a schematic
representation of 2DA as well as an AFM image of the chess board fields.
$\phi$ scans reveal a predominant $[100]$ CeO$_2$ $\parallel$
$[100]$ MgO growth of the buffer layer, but with a fraction of $[110]$
CeO$_2$ $\parallel$$[100]$ MgO orientations. In addition, some small
fraction of grains having a mutual misorientation angle of $\pm$ 24$^o$
was detected. From this it is clear, since LSMO inherits the
orientaion of CeO$_2$, that GBF also will contain
GB of the kind indicated by the $\phi$ scans.\\
 There is limited data on the structure of GB in
LSMO. We have performed detailed TEM studies on a 20$^o$ GB grown on
a LaAlO$_3$ bicrystal substrate\cite{zdravkap2}. The two parts of the LSMO
film on
the bicrystal
substrate form a sharp on an atomic scale symmetrical GB. However, facets
parallel to the
low index crystallographic planes of the LSMO are present. The GB
consists of closely spaced edge dislocations with a periodicity of 4-5
LSMO unit cells (1.6-2 nm). No impurity phases are detected at the GB.
Although we have not investigated the 45$^o$ GB in 2DA, from the results on
the 20$^o$ GB, we expect the 45$^o$ GB to have a similar structure
but with even closer spaced edge dislocations. The disorder at the GB can
be estimated
to have a thickness of 3-5 nm in the LSMO layer.\\
Magnetization $M(H,T)$ measurements were performed in a Quantum Design
SQUID magnetometer. The resistivity $\rho(H,T,\theta)$
was measured using a standard four-probe method and a Maglab 2000
system
from Oxford Instruments with a rotationary probe. The
magnetoresistance of
the samples is defined as $(R_0-R_H(\theta))/R_0$; the angle $\theta$
refers to
the angle between the current and the in-plane applied magnetic field.

\section{Results from magnetic and transport measurements}

Fig.~\ref{fig1} shows the temperature dependence of the magnetization
for all samples. Zero-field cooled
(ZFC) and field cooled (FC) magnetizations with a magnetic field of 4
kA/m are shown. All films exhibit ferromagnetic order at low temperature
with approximately the same Curie
temperature $T_c \approx$ 360 K, in agreement with results from
earlier studies
on LSMO films with optimum hole doping\cite{rami}. The low field
magnetization below $T_c$ is larger for EF than for the
other two samples. This is expected, considering the high crystalline
quality of this sample, since reversible and irreversible
domain wall motions determine the magnitude of the low field
magnetization. For 2DA, the
magnetization remains large and approximately
constant above $T_c$, indicating some kind of magnetic order
remaining in the
sample even at these high temperatures, a conclusion which is further
supported by the hysteresis curve shown in the inset of
Fig.~\ref{fig1}. X-Ray diffraction reveals no impurity phases,
suggesting a real two-step magnetic transition. This peculiarity of the 2DA
sample is not fully understood; it could be related to the specific
properties of this kind of grain boundary. The previously discussed
study of a 20$^o$ bi-crystal GB revealed a
regular set of edge dislocations with a period of 4-5 unit cells and
strong stress fields at the grain boundary\cite{zdravkap2}. These
two factors may contribute to the observed high temperature magnetic ordering.
The origin of this ordering is however left for further studies where
the size of the chess board fields will be varied, thereby changing the
relative amount of distorted film material. \\
The field dependence of the magnetization was studied at different
temperatures in the range 5 K to 200 K. Typical hysteresis curves are
shown in Fig.~\ref{fig2} for $T$=5 K. For EF, the hysteresis curve is
rather square shaped, as in a
sample with no (or few) defects, confirming the excellent epitaxial
growth. GBF contains some amount of grain boundaries; as a result, the
hysteresis curve is more inclinated. Also one notices that the addition of
defects in the form of grain boundaries promote the nucleation of
reversed domains, thereby reducing the coercivity. Adding more boundaries
as is the case for the chess
board film, the hysteresis curve becomes even more
inclinated, but the coercivity increases, indicating a pinning
controlled mechanism for the coercivity in this sample. These general
characteristics remain at higher temperatures.
Fig.~\ref{fig3} presents the zero magnetic field resistivity of the
two bi-epitaxial films, with the results for EF as an inset for
comparison. The behavior of the bi-epitaxial films is very different
from that of EF, with no significant features at $T_c$. Broad maxima in
the resistivity are present well below $T_c$, like in the resistivity
curves obtained using a Wheatstone bridge geometry on
La$_{0.7}$Sr$_{0.3}$MnO$_3$ bi-crystal meander-patterned films to measure
directly the grain boundary
resistivity\cite{evetts}, indicating a grain boundary dominated
resistivity for the two bi-epitaxial films.\\
Fig.~\ref{fig4} shows the temperature
dependence of the low field magnetoresistance (for $\mu_0H$= 0.1
Tesla) for the two bi-epitaxial films, with the corresponding result for
EF as an inset in the same figure. EF displays the typical low field
magnetoresistance behavior for a high quality epitaxial
film\cite{hund1,donnell1,steenbeck12,shreekala,ziese12,epit1,epit2},
with a peak in the
magnetoresistance around $T_c$, and no significant low temperature
MR. To observe a ``colossal'' magnetoresistance in this sample, much
larger fields are needed; the MR is 10 \% at RT applying a field of $5$ T.
These results are typical for single crystals and high quality epitaxial
films of LSMO\cite{evetts}.
Due to the strong transport-magnetism correlations seen in high
quality CMR films, the conduction is thought to correspond to activated
(magnetic) polaron hopping\cite{helmont,hund1,donnell1,millis}, even
though other mechanisms like reduction of spin fluctuations have been
suggested to account for the MR effect\cite{hwang1,furukawa}. In comparison,
the bi-epitaxial films exhibit a low
temperature tail with an increasing low field magnetoresistance with
decreasing temperature. No significant features appear at $T_c$. On the
one hand, the absence of a magnetoresistance peak around $T_c$ can be
attributed to grain boundary stress fields and/or stoichiometry
variations\cite{evetts},
which will change the Curie temperature close to the grain boundary or
may even locally create a different type of magnetic order. It is to
be expected that the different types of grain boundaries existing in
the bi-epitaxial films are
associated with distributions of stress fields (and stoichiometry
variations) and hence distributions of
grain boundary Curie temperatures, thereby erasing the sharp
magnetoresistance peak around the measured $T_c$. On the other hand, the
low temperature raise of the magnetoresistance is attributed to a different
transport mechanism such as spin polarized tunneling
through a barrier region, something which will be discussed in
more detail below. \\
Fig.~\ref{fig56} (a) shows the high field behavior of the
magnetoresistance for 2DA at different temperatures, from $T$=10K to
$T$=300K,  and Fig.~\ref{fig56} (b) the low-field behavior for
different orientations of the in plane magnetic field ($H \parallel I$ and $H
\perp I$); similar features were observed for the GBF film. The high
field resistance at first sight
looks linear with the magnetic field (cf. Fig.~\ref{fig56} (a)) ,
but, as will be
discussed later, it is the high field conductance that exhibits a linear
high field regime. 2DA also shows magnetoresistance hysteresis at low
fields (cf. Fig.~\ref{fig56} (b)), commonly related to defects and grain
boundaries in the films; the peak resistance
occurs at a field near to the coercive field. The hysteretic
behavior remains at higher temperatures. One also notices that, as
has been reported for structures with well oriented grain boundaries
\cite{evetts},
a higher MR effect is obtained for $H \parallel I$.\\
If the previously discussed features were expected, considering the
presence of grain boundaries, peculiar  orientation-dependent effects appear
for the bi-epitaxial films. In Fig.~\ref{fig7}, resistivity vs. $\theta$,
angle between the applied magnetic field and current, is presented for
the two bi-epitaxial films at
$T$=80 K and $\mu_0 H$=0.05 T (Fig.~\ref{fig7}(a) and (b)). Both
samples exhibit anisotropic
magnetoresistance (AMR). Sinusoidal $\alpha_2$sin(2$\theta$) fits are
included; the residue subtracting the fit from the experimental
result is
shown in
Fig.~\ref{fig7}(c) and (d). For 2DA, a new sinusoidal fit of the
residue
has been added, suggesting an additional $\alpha_4$sin(4$\theta$)
periodic contribution. This term disappears when increasing the
magnetic
field above the saturation field of the sample (Fig.~\ref{fig8}), at
which fields the AMR amplitude also saturates; the
high field AMR is $\approx$1.5 \%. The residue
for GBF is smaller, but still suggests contributions from higher
frequency
angular terms. Fourier analysis of the angular dependence of the
magnetoresistance allows us to resolve (at least) $2\theta$-,
$4\theta$- and $6\theta$-terms; for the GBF film $\alpha_{2} >>
\alpha_{i
> 2}$, while for the 2DA film $\alpha_{4}$ and $\alpha_{6}$ are
comparably large (approximately 1/10 of $\alpha_{2}$). The AMR in EF
is
rather much
smaller as compared to that displayed in Fig.~\ref{fig7}; the low
temperature, high field AMR is only about 0.2-0.3 \%. Still, a Fourier
analysis of the angular dependence of the magnetoresistance for this
film
shows a behavior similar to that of 2DA; at low fields $2\theta$-,
$4\theta$- and $6\theta$-terms can be resolved while for fields larger
than the saturation field only the $2\theta$-term is seen. One the one
hand, the high field AMR is an intrinsic
property of LSMO associated with spin-orbit
coupling\cite{ziese1,sinus}. On
the other hand, the low field AMR contains an extrinsic contribution
from the
geometry of the grain boundaries as well as a contribution originating
from, and having the same symmetry as, the magnetic anisotropy.
Below but close to saturation, the induced magnetization will be modulated
as determined by the symmetry of the magnetic anisotropy when the film is
rotated with respect to the applied field.\\

\section{discussion}

It is clear that one additional transport mechanism is present in films
containing grain boundaries as compared to high quality epitaxial films.
A model attempting to describe the properties of the bi-epitaxial
films must be able to account for; i) the low temperature tail of the low
field
magnetoresistance; ii) the high field behavior of the
magnetoresistance (or the magnetoconductance); and iii) the AMR
behavior.\\
A model including spin polarized tunneling best explains our experimental
results. A tunneling junction can be modelled as a resistor\cite{mahan},
with the resistance
given by $R_j=1/G_j$, where $G_j$ is the tunneling conductance. The basic
building block in our bi-epitaxial films is therefore $R=R_j+R_e$, where
$R_e$ is the resistance of the LSMO ferromagnetic electrodes. For
temperatures $T \ll T_c$,  $R_j \gg R_e$ holds even though $R_e$ of
the bi-epitaxial film due to lattice strain may be larger than the
resistivity of the EF film.
\\
Magnetoresistance measurements on single magnetic tunnel junctions in
general show step like features between high and low resistance
states of the junction at fields corresponding to the coercive field
of the structure\cite{li,sun}. For our grain boundary samples, the $\rho(H)$
curves exhibit less sharp features (confer Fig.~\ref{fig56} (b)), which
is an effect caused by dispersion in the parameters controlling the
spin polarized tunneling process.\\
In the original work of Julliere\cite{julliere}, an assumption of spin
conservation in the tunneling process was made and the
magnetoresistance
was simply expressed in terms of the spin polarizations $P_{1,2}$ of the
two ferromagnetic electrodes;
$P=(n_{\uparrow}*-n_{\downarrow}*)/(n_{\uparrow}*+n_{\downarrow}*)$, where
$n_{\uparrow}*$ and $ n_{\downarrow}*$ are the electronic density of states
for
majority and minority carriers, respectively.
To explain the observed temperature
dependence of the magnetoresistance for the films containing grain
boundaries, it is necessary to add to
the spin conserving tunneling process the possibility of spin-flip
tunneling\cite{gu,guinea,lyu1,lyu2,inoue}, eg. induced by magnetic
impurity states inside the barrier\cite{gu} or by
spin wave excitations at the barrier surface\cite{lyu2}. Another
possibility of explaining this temperature dependence is linked to the
intrinsic spin polarization\cite{lyu2} in CMR materials. At low
temperature,
experimental results indicate half-metallic behavior\cite{okimoto,park,wei},
i.e. complete spin polarization ($P_{1,2}$=1), results which are
corroborated by
band structure calculations\cite{pickett,satpathy}. However, the experimental
results \cite{okimoto,park,wei} also indicate that the electronic structure
varies
with temperature. This led Lyu et al.\cite{lyu2} to propose a model
where the temperature dependent tunneling magnetoresistance is an
effect resulting from a temperature dependent spin polarization in
combination with
collective spin excitations at interfaces. It was also shown
that this model could reproduce the main features of the temperature dependent
magnetoresistance obtained for a LSMO/STO/LSMO trilayer junction\cite{li}.
The similarity between the results shown in Fig.~\ref{fig4} for the
temperature dependent magnetoresistance and the results obtained
for the trilayer junction suggests that the model of Lyu et al.\cite{lyu2}
also
applies for the GBF and 2DA samples. More specifically, the model
correctly predicts a strong decrease of the magnetoresistance at a temperature
much lower than $T_{c}*$. \\
To be able to account for the high field behavior of the magnetoresistance
it is necessary to consider the magnetic properties of the grain boundary
itself. Here it should be pointed out that the
observed slope of the high field conductance is much larger than that
observed for the epitaxial film, and therefore it is not possible to
assign this high field behavior to the LSMO electrodes. A linear high field
regime for the conductance has previously been
reported by Lee et al., who studied the magnetotransport behavior of
polycrystalline manganite samples\cite{lee}.
In the same paper, it was shown that the experimental results were
consistent with an interpretation based on second-order tunneling
through
interfacial spin sites. Using the transfer
integral
$T_{12} \propto \sqrt{1+\stackrel{\rightarrow}{s_1} \cdot
\stackrel{\rightarrow}{s_2}}$
for intinerant $e_g$ electrons between localized $t_{2g}$
moments ($\stackrel{\rightarrow}{s_1}$ and
$\stackrel{\rightarrow}{s_2}$
are the normalized spin moments), the conductivity $G_j$ was given as,
\begin{equation}
G_j \sim {T_{1j}}^2{T_{j2}}^2 = \langle
(1+\stackrel{\rightarrow}{s_1}\cdot \stackrel{\rightarrow}{s_j})\cdot
(1+\stackrel{\rightarrow}{s_j}\cdot
\stackrel{\rightarrow}{s_2})\rangle
\label{eq1}
\end{equation}
where $\stackrel{\rightarrow}{s_j}$ is the normalized grain boundary
spin
moment and $\langle \ldots \rangle$ denotes thermal average. For
large
enough field, having saturated the magnetization of the two LSMO
electrodes, and to lowest order in field, one obtains
$G_j \sim \langle \stackrel{\rightarrow}{s_j} \rangle \propto
\chi_jH$, where $\chi_j$ is the susceptibility of the boundary region.
Our own results for the bi-epitaxial films also show that the
magnetoconductance, rather than the magnetoresistance, exhibits a
linear high field regime. The initial conductivity rise (before the
linear
regime) at low temperature is close to 30 \% for 2DA (confer Fig. 5
(a)),
in agreement with
the upper limit of 33 \% predicted by the model.\\
The magnetic properties of the boundaries as
given by the temperature dependence of the normalized high field
slope of
the magnetoconductance $b(T)=dG/\mu_0G_0dH \propto \chi_j$ is shown in
Fig.~\ref{fig9}. The results obtained by
Lee et al.\cite{lee} for a polycrystalline
La$_{0.67}$Sr$_{0.33}$MnO$_3$
sample are included for comparison. It is noteworthy that the
properties of the grain boundaries are so similar in the bi-epitaxial
films and in bulk polycrystalline LSMO samples, indicating that the
magnetism close to an interface is determined by intrinsic rather than
extrinsic properties. The temperature dependence of
the high field $\chi_j$, with a weak increase with decreasing
temperature, suggests some
kind of disordered magnetic state in the grain boundary region. As to the
true nature of this state, it is not possible to give a definite
answer only on the basis of the present study. In passing, we note
that a different model, not based on spin polarized tunneling, has been
proposed by Evetts et al.\cite{evetts} to describe the observed
magnetoresistance
behavior of artificial grain boundaries in thin film bi-crystals. This
model depends on activated transport within grain boundary regions,
and the magnetoresistive response is determined by the grain
boundary magnetization. While this model is capable of explaining some
of the features observed for the bi-epitaxial films, it predicts a
linear high-field regime for the resistance rather than for the conductance. \\
The model, as formulated by Lee et al.\cite{lee}, does not contain an
anisotropic term to relate to the angular profiles of the resistivity
shown in Fig.~\ref{fig7}. Ziese and Sena\cite{ziese1} developed an atomic
model to explain the AMR in CMR
materials. In this model the AMR amplitude is expressed in terms of
intrinsic local parameters like the spin-orbit coupling, the crystal-
field and exchange-field splittings. This implies that both the
electrode and the grain boundary near regions
exhibit an AMR effect of the same atomic
origin. The larger AMR observed for the bi-epitaxial films can be
attributed to stress fields associated with the grain boundaries. The
resulting strain may change intrinsic properties such as the
crystal-field splitting locally, thereby affecting the AMR
amplitude\cite{ziese1}. \\
Below saturation, the AMR of EF contains $4\theta$-
and $6\theta$-terms comparably large in magnitude, something which
can be attributed to the symmetry of the magnetic anisotropy. These higher
frequency angular terms are rather much smaller in magnitude for GBF,
while for 2DA the relative magnitudes of these terms again are large. It is worth noting that this observation cannot
be explained by differences in domain configurations and possible domain wall
contributions to the magnetoresistance, since the
results discussed here correspond to the reversible or near to reversible
magnetization regime with the applied field being much larger than the
coercive field. The reinforcement of the $4\theta$- and $6\theta$-terms in
case of 2DA is instead attributed to the existence of oriented grain
boundaries
in this film. To account for this, an anisotropic term with the same
symmetry as the grain boundary array is included by replacing, as
suggested by Evetts et al.\cite{evetts,evetts3}, the applied field
with the local field $H_j$ acting on the grain boundary region,
\begin{equation}
\langle\stackrel{\rightarrow}{s_j}\rangle = \chi_{j} \cdot
(H+f(\phi)M_e)
\label{eq3}
\end{equation}
where $f(\phi)$ is a geometric factor and $M_e$ the saturation
magnetization of the LSMO electrode.
This additional term does not contribute significantly at high
fields, so
the linear high field behavior of the conductivity is
preserved, but adds an orientation dependent term to the conductivity,
$f(\phi)$ depending on the orientation of applied field with respect to
the grain boundary array. Thus, the reinforcement
of the higher frequency angular terms observed for FEBC is a result of
creating an artificial square array of grain boundaries in this
sample.

\section{Conclusion}

We have compared the magnetic and transport properties of bi-epitaxial
films of
La$_{0.7}$Sr$_{0.3}$MnO$_3$ with the corresponding properties of a
high
quality epitaxial film. Both
bi-epitaxial samples exhibit a grain boundary dominated resistivity,
and
the magnetoresistance results are well described by a two-step spin
polarized tunneling mechanism. Additional anisotropic
magnetoresistance
effects are discussed, and found to have both intrinsic (magnetic
anisotropy) and extrinsic (grain boundary distribution) origins. The
two-step
tunneling model originally proposed by Lee et al.\cite{lee} is
modified to
include the anisotropic features. Surprinsingly, for the 2D array, a
constant magnetization was observed above the Curie temperature of
LSMO,
indicating a magnetic ordering of unknown origin.

\acknowledgements
This work was financially supported by The Swedish Natural Science
Research Council (NFR).

\pagebreak
\begin{figure}
\caption{An AFM image of 2DA and a schematic representation
of the orientation of the different layers of this bi-epitaxial
structure.}
\label{fig0}
\end{figure}

\begin{figure}
\centerline{\epsfig{figure=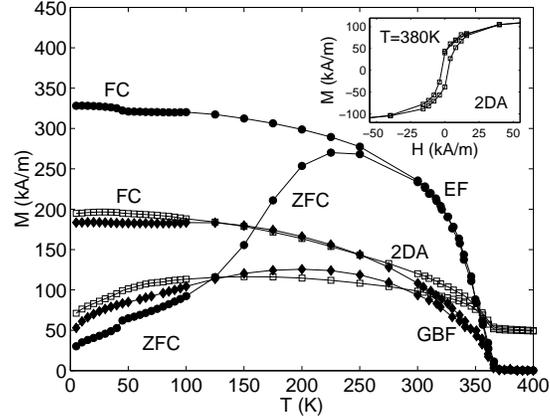,height=6cm,width=8cm}}
\caption{ZFC and FC magnetization for EF (filled circles), GBF (filled
diamonds) and 2DA (open squares); $H$=4 kA/m. The inset shows the
hysteresis loop for 2DA at T=380K.}
\label{fig1}
\end{figure}

\begin{figure}
\centerline{\epsfig{figure=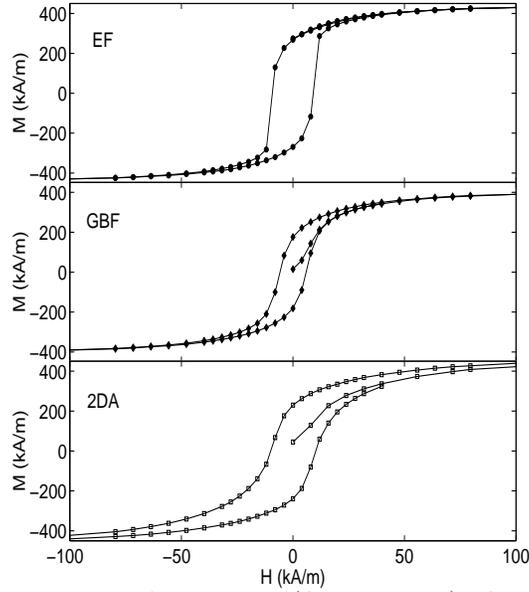,height=8cm,width=7cm}}
\caption{Hysteresis loops at T=5K for the EF (filled circles), GBF
(filled diamonds) and 2DA (open squares) samples.}
\label{fig2}
\end{figure}

\begin{figure}
\centerline{\epsfig{figure=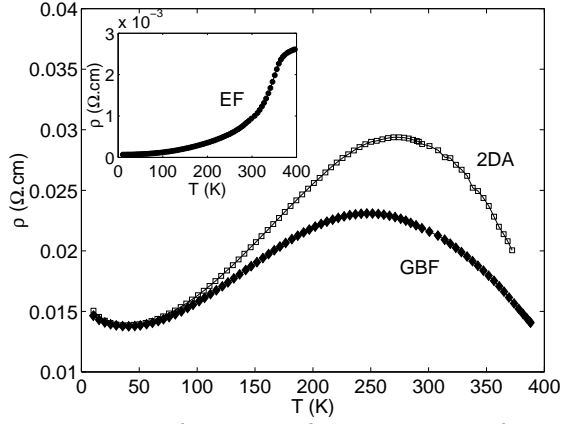,height=6cm,width=8cm}}
\caption{Temperature dependence of the zero field resistivity for the two
bi-epitaxial films. The inset shows the corresponding result for EF.}
\label{fig3}
\end{figure}

\begin{figure}
\centerline{\epsfig{figure=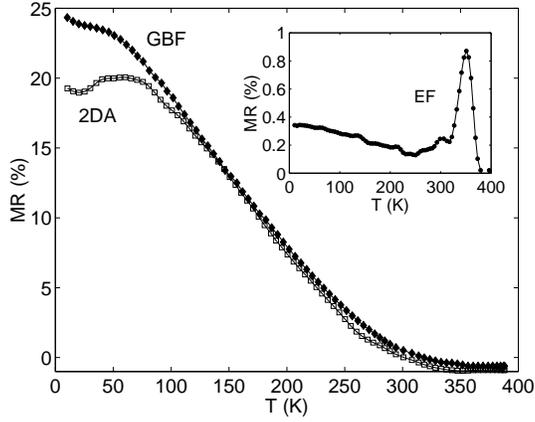,height=6cm,width=8cm}}
\vspace*{0.5cm}
\caption{Temperature dependence of the low field magnetoresistance
($\mu_0 H$=0.1 T) for the two bi-epitaxial films. The inset shows the
corresponding result for EF.}
\label{fig4}
\end{figure}

\begin{figure}
\centerline{\epsfig{figure=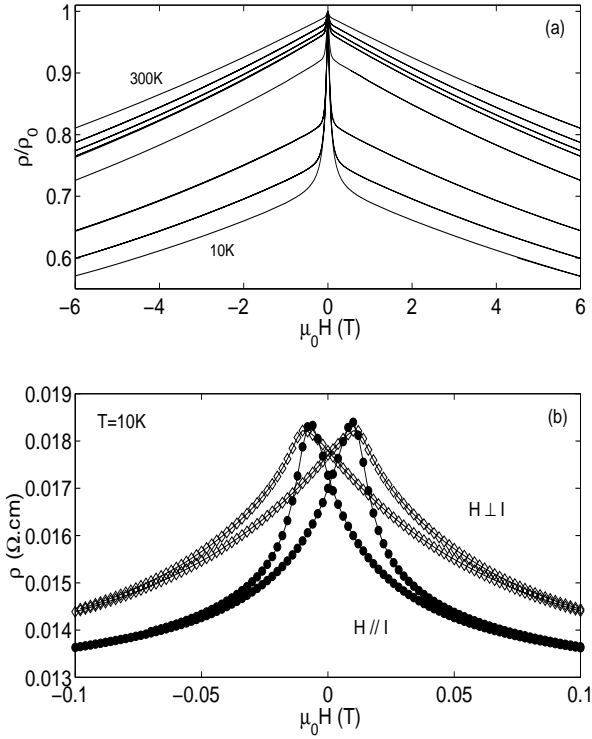,height=10cm,width=8cm}}
\vspace*{0.5cm}
\caption{Field dependence of the resistivity of 2DA for high fields
(a) and
low fields (b). For high fields, the normalized resistivity is shown
for
$T$=10, 50, 100, 200, 250, 260, 275, and 300 K; for clarity, only 10K
and
300K are marked in the figure. In the low field case, hysteresis loops
are shown for $T$=10 K with the applied magnetic field parallel and
perpendicular to the current.}
\label{fig56}
\end{figure}

\begin{figure}
\centerline{\epsfig{figure=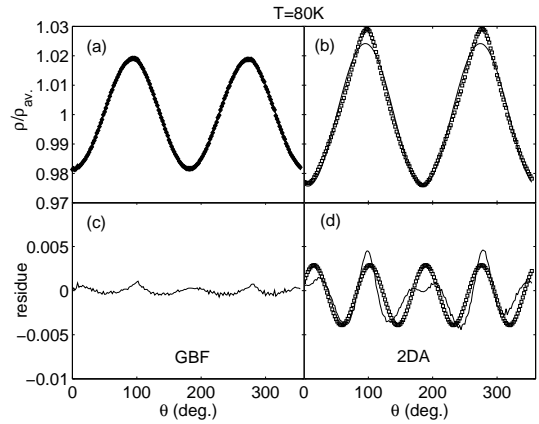,height=6cm,width=8cm}}
\vspace*{0.5cm}
\caption{Angular dependence of the resistivity for the two
bi-epitaxial
films; $\mu_0H$= 0.05 Tesla.
The normalized resistivity is fitted by a sinuisoidal function ((a)
and (b)),
and the residue subtracting the fit from the experimental data is
shown in
(c) and (d). In (d) an additional sinusoidal fit of the residue is
included.}
\label{fig7}
\end{figure}

\begin{figure}
\centerline{\epsfig{figure=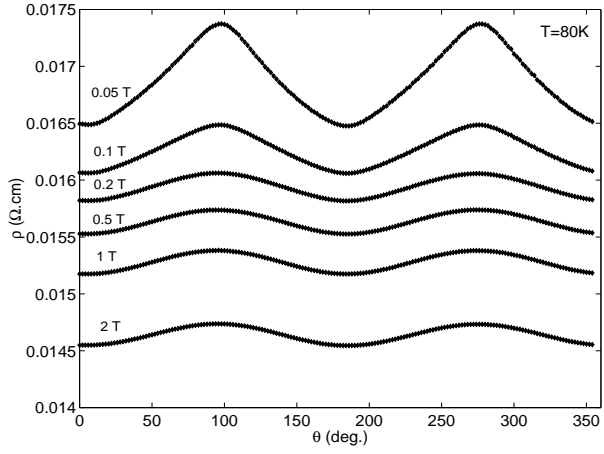,height=6cm,width=8cm}}
\vspace*{0.5cm}
\caption{Field dependence of the AMR for 2DA.}
\label{fig8}
\end{figure}

\begin{figure}
\centerline{\epsfig{figure=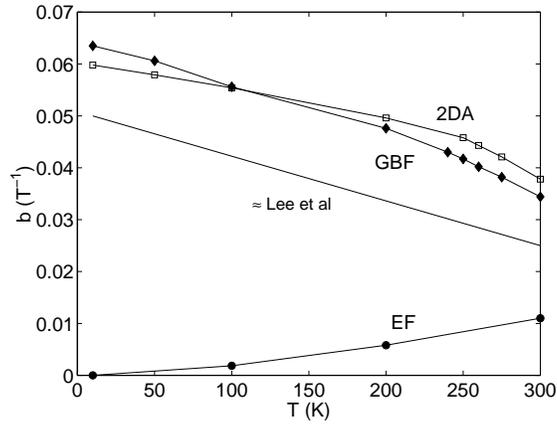,height=6cm,width=8cm}}
\vspace*{0.5cm}
\caption{Temperature dependence of the normalized high field
magnetoconductance slope $b(T)=dG/\mu_0G_0dH$; Earlier results by
Lee et al.[\protect\onlinecite{lee}] are included for comparison
(solid line).}
\label{fig9}
\end{figure}

\end{document}